# The Localized 12-hour Wave Over Alaska: Leveraging Meridional Wind Measurements From the Sodium Lidar

Sophie R. Phillips[1], Katrina Bossert[1], Komal Kumari[2], Erich Becker[3], Richard L. Collins[4], Nicholas Pedatella[5]

[1]Arizona State University, Tempe, AZ, USA

[2]University of Illinois – Urbana-Champaign, Champaign, IL, USA

[3]NorthWest Research Associates, Boulder, CO, USA

[4]University of Alaska, Fairbanks, Fairbanks, AK, USA

[5] National Center for Atmospheric Research, Boulder, CO, USA

Corresponding author: Sophie R. Phillips (sphillips@asu.edu)

## Key Points:

- Large 12-h wave amplitudes (up to 160 m/s at 102 km) were observed in lidar meridional winds during the 2018-2019 season
- Meteor radar measurements of the meridional 12-h wave amplitudes demonstrate underestimates compared to lidar measurements
- Large 12-h wave amplitudes were not linked to strong geomagnetic activity, and occurred during a sudden stratospheric warming

## Abstract

The 12-h wave in meridional winds in the mesosphere and lower thermosphere (MLT) during the solar minimum 2018-2019 Arctic winter is investigated using sodium lidar observations at Poker Flat Research Range (PFRR), Chatanika, Alaska (64°N,147°W). Nightly 12-h wave amplitudes increased significantly during December-January, with amplitudes exceeding 130 m/s above 97 km on several days. This was more than double the 12-h wave amplitudes observed outside this time period. Meteor radar winds over Chatanika also showed significant increase in meridional wind 12-h wave amplitudes at altitudes between 82 and 97 km during this time period with lower amplitudes than lidar measurements. The strong variation in 12-h wave amplitudes was not correlated with SME index, though the largest amplitudes coincided with the sudden stratospheric warming (SSW) in early January. Measurements were compared to a seasonal WACCM-X model run and four days of HIAMCM. For the four dates of available HIAMCM data, 12-h wave amplitudes over Chatanika were found to be similar between both HIAMCM and WACCM-X and the lidar below 97 km, with amplitudes measured by lidar exceeding the models at altitudes between 97-105 km. All measurements followed a similar seasonal trend with increasing amplitudes at the end of December/early January. Fits of SW2 from WACCM-X show the SW2 tidal amplitude following similar seasonal trends to 12-h wave measurements. These high-resolution lidar measurements indicate that localized 12-h wave amplitudes are larger than previously reported by studies using meteor radar measurements.

## 1 Introduction

The Mesosphere, Lower Thermosphere (MLT) is a dynamic region that couples the middle atmosphere with the ionosphere and thermosphere. Atmospheric tides are one of the primary drivers of dynamics in this region (Lieberman et al., 2026). The main tidal components are diurnal (24-h), semidiurnal (12-h), and terdiurnal (8-h), with the 12-h wave dominating in the Arctic MLT (Ramesh et al., 2025). In particular, the 12-h tide's amplitude reaches daily localized amplitudes of up to 40 m/s at ~95 km when measured by meteor radar (Dempsy et al., 2021) and peaks between 100-111 km (Oberheide et al., 2011). Tides experience day-to-day variability due to nonlinear interactions with planetary waves and GWs (Dempsy et al., 2021), and therefore significantly affect short-term variability in the upper atmosphere (Dhadly et al., 2023). Currently, there are limited measurements in the polar region, yet studies have suggested that the SW2 tide is significant at these high latitudes (Ramesh et al., 2025; Chau et al., 2015; Goncharenko et al., 2023).

Gravity Waves (GWs) also play a significant role in the structure of the MLT. GWs are generated by many sources, such as flow over orography (Bossert et al., 2017; Heale et al., 2020; Heale et al., 2022; Holt et al., 2017; Becker et al., 2025), convection (Hoffman & Alexander, 2010; Heale et al., 2022; Vadas & Liu, 2013; Hoffmann et al., 2018; Xu et al., 2019), as well as spontaneous emission from tropospheric jets and fronts and the polar night jet (Plougonven & Zhang, 2014; Sato & Yoshiki, 2008; Bossert et al., 2020; Becker et al., 2022a; Vadas et al., 2024). They deposit energy and momentum upon breaking, which creates localized body forces that both accelerate the mean flow as well as give rise to higher order GWs (Vadas et al., 2003; 2018).

Global coverage of winds in the MLT is limited, though studies indicate that winds in this region can vary significantly. Larson (2002) presented results of strong wind variability over decades of rocket launches. Wind speeds tend to peak between 100-110 km with magnitudes commonly exceeding 100 m/s (Figures 9 and 10 from Larson, 2002), and these peak altitudes have an associated large shear below the maximum altitudes. Additionally, high latitude vertical wind speeds can reach amplitudes up to 20 m/s in the lower thermosphere (Larson & Meriwether, 2012; Meriwether et al., 2026). Due to these strong winds, significant vertical shears have been observed between 100-130 km, and observed meridional shears at these altitudes have been slightly stronger than zonal shears (England et al., 2022).

Conditions in the stratosphere are known to affect the thermosphere region, including impacts on GWs (Heale et al., 2020), tides (Kumari et al., 2021; Oberheide et al., 2009), and ionospheric density variations (Goncharenko et al., 2013). The main aspect of the stratosphere in this respect is the variability of the polar vortex. Kumari et al. (2023) showed that at high latitudes, while Traveling Atmospheric Disturbance (TAD) activity in the thermosphere decreases correspond with sudden stratospheric warmings (SSWs), times of strong thermosopheric TADs correlated with positive changes in stratospheric planetary wave amplitudes. In the F region ionosphere, Medium Scale Traveling Ionospheric Disturbance (MSTID) activity also decreases during SSWs, though more substantially at mid-latitudes (Frissell et al., 2016; Becker et al., 2022b). Phillips et al. (2026) showed a 43% decrease in MSTID amplitudes over Alaska during an SSW and suggested MSTIDs are further influenced by winds at 50 km in addition to polar vortex strength. High-latitude, large-scale wind variability was reported in the MLT during a rare southern hemisphere SSW (Liu et al., 2022).

Kumari et al. (2025) showed that zonal winds associated with the semidiurnal tide at high latitudes doubled from the MLT to the upper thermosphere during an SSW while geomagnetic activity was low. The study presented here expands upon this work by investigating meridional winds for the same time period investigated by Kumari et al. (2025), namely, the Arctic 2018-2019 winter. By analyzing high-resolution lidar data, we will provide further insight into localized 12-h wave amplitudes during this time period over Alaska.

This study presents observations and model results of 12-h wave meridional wind amplitudes observed over Alaska during solar minimum. Section 2 describes the instruments and models used in this research, as well as the techniques used to isolate the 12-h wave. Section 3 compares the results of the 12-h wave amplitudes from the observations and models. We provide a discussion of the results and how they compare to previous studies in Section 4. Section 5 summarizes our results and suggests future work.

## 2 Instrumentation and Methods

### 2.1 Sodium Lidar

The sodium lidar used in this work is located at the PFRR Lidar Research Laboratory. This sodium lidar retrieves sodium density, temperature, and meridional wind measurements between ~80-110 km during clear night conditions (Li et al., 2020). Raw lidar signal profiles are recorded every 7 seconds for this study. These profiles are integrated over a period of 20 minutes and an altitude interval of 1 km to reduce noise, especially at higher altitudes. The sodium lidar was configured with two beams; a zenith beam and a north beam pointing 20° off-zenith. The zenith beam yielded measurements of sodium density and temperature, while the north beam yielded measurements of sodium density, temperature, and meridional wind. No zonal winds were measured with the configuration used in this study. For the 2018-2019 Arctic winter, nighttime data was acquired on 13 occasions: 9 October 2018, 19 October 2018, 2 November 2018, 6 November 2018, 28 November 2018, 23 December 2018, 24 December 2018, 5 January 2019, 7

January 2019, 18 January 2019, 2 February 2019, 15 February 2019, and 2 March 2019. Altitude ranges where errors >10 m/s persisted for extended periods were excluded.

In order to isolate the 12-h wave in the sodium lidar data, the following harmonic function is fitted to the meridional winds for each night:

$$a(t)=A*\sin((2\pi/P)*t) + B*\cos((2\pi/P)*t) + m. \tag{1}$$

Here, P is the period of the wave (12 hours) and m is the mean wind over the observation. The wave amplitude is calculated as the RMS amplitude, $\sqrt{A^2 + B^2}$. This is done at each altitude. The lidar operates at nighttime, which is extended in the polar winter region but does not extend over the 24-h period at Chatanika. The data is shown with respect to UT time. The lidar observation periods ranged from 8.4 hours (6 November 2018) to 14.8 hours (5 January 2019), with an average of 11.6 h for these 13 nights. The lidar operated on 16 January 2019 for 4.8 h, which was deemed insufficient for calculating a 12-h wave amplitude. The fitting method is applied for each of the 13 days of available lidar data in the 2018-2019 Arctic winter at each altitude bin. Each meridional wind speed value has a corresponding uncertainty. In fitting for the 12-h curve, uncertainties at each altitude are calculated by averaging wind uncertainties over the entire night.

### 2.2 Meteor Radar

The meteor radar used in this study is also located at PFRR (65.1° N, 147.5° W). Using ionized particle trails from burned meteors embedded in the background wind, it continually derives zonal and meridional winds (Klemm, 2019). Data is therefore available for both day and night between 82-98 km, with 3-km vertical resolution and 1-h temporal resolution. In ensuring data quality, meteor echoes for which the calculated zenith angle is over 85° are rejected. Therefore, the radar's field of view expands ~170 degrees.

A similar method as was used in the sodium lidar was applied to radar measured meridional winds to obtain the 12-h wave. Since the radar provides continuous data, the curve fitting function was applied over 1-day intervals. The wave amplitude is calculated as before, using $\sqrt{A^2 + B^2}$. This process is done for the entire season of data, starting with the radar's installation in November 2018, up to March 2019.

**2.3 WACCM-X**

The Whole Atmosphere Community Climate Model with thermosphere-ionosphere eXtension (WACCM-X) is a numerical atmospheric model that simulates the atmosphere from Earth's surface to the upper thermosphere (Liu et al., 2010; Liu et al., 2018). Our study presents results from a high-resolution configuration, described in Liu et al. (2024). The simulations used in the present study are described in Pedatella et al. (2026) and include constraining the model meteorology to ERA-5 reanalysis. This simulation includes parameterized GWs generated by convection, fronts, and orography to account for GWs that are unresolved at ~25 km horizontal resolutions (horizontal wavelengths $\lambda_H < \sim 200$ km). High resolution WACCM-X has been shown to improve the simulations of upper atmospheric dynamics during SSWs, such as zonal winds and nitric oxide (NO) transport, when compared to lower resolution runs (Pedatella et al., 2026). Output includes meridional winds for different locations at a 1-h temporal resolution, 0.25° latitudinal and 0.25° longitudinal spatial resolution. In the MLT, the model vertical resolution is 0.1 scale height (ranges from 0.7-1 km in the MLT region). Our data-model validation will focus on the specific location of PFRR.

The results presented from WACCM-X mainly include two data outputs: 12-h fit meridional amplitudes over PFRR, and zonally derived migrating semidiurnal tide (SW2) meridional wind amplitudes. The SW2 amplitudes are calculated by taking a 2D FFT along longitude and UT, over a 3-day moving window to account for the correct tidal phase. The 12-h fits use a 1D FFT along UT. While the SW2 is a more accurate measure of the global tidal activity, the 12-h wave fit is also used for more direct comparison with the stationary lidar and radar data sets.

**2.4 HIAMCM**

The HIgh Altitude Mechanistic general Circulation Model (HIAMCM) is a global atmospheric circulation model that spans from Earth's surface up to ~450 km (Becker & Vadas, 2020). The model employs a spectral dynamical core (Simmons & Burridge, 1981) with high horizontal and vertical resolution as well as a physics-based subgrid-scale diffusion such that medium scale

GWs down to horizontal wavelengths of ~200 km are simulated explicitly, including secondary and higher-order GWs resulting from the localized and intermittent dissipation of GWs from farther below (Becker & Vadas, 2020; Vadas et al., 2024; Becker et al., 2025). Importantly, the HIAMCM does not include a GW parameterization. For the results presented in this work, the model was run with nudging to MERRA-2 reanalysis, enabling one-to-one comparison with observations. To preserve the self-consistent simulation of GWs, the nudging is performed in spectral space and applied to the large scales only (Becker et al., 2022a).

## 3 Results

The 20-minute temporally, 1-km vertically averaged meridional winds for three select nights from the sodium lidar are shown in Figure 1a-c (top row). From this higher resolution lidar data, a moving average over 1-h temporal and 3-km vertical resolution is performed for better comparison with the meteor radar winds. The 20-minute, 1-km resolution lidar data are compared to HIAMCM (second row), 1-h and 3-km averaged lidar data (third row), meteor radar (fourth row), and WACCM-X (bottom row) meridional winds over PFRR.

The sodium lidar measures strong meridional wind amplitudes from 83-105 km, which generally grow larger at higher altitudes. 18 January shows the largest amplitudes near 97 km, and 23 and 24 December show largest amplitudes above 100 km. When comparing the lidar data to HIAMCM (Figure 1d-f), the meridional winds are similar, but slightly lower in magnitude, and both lidar and HIAMCM demonstrate agreement in phase structure of the 12-h wave.

Figure 1g-i in the third row shows the averaged lidar data with a 1-h and 3 km smoothing. After averaging, the lidar shows a similar wind structure and wave phases to the meteor radar measurements (Figure 1j-l). Radar meridional wind measurements are overall lower in magnitude than the meridional winds from the averaged lidar data.

The meridional wind variations from WACCM-X (Figure 1m-o) are on the same order of magnitude as the wind speeds from the lidar and radar. For 23 December and 24 December, the 12-h wave is present in the WACCM-X data with comparable, yet slightly lower, wind amplitudes. On 18 January, the WACCM-X meridional wind speeds are notably lower than the lidar, especially to the upper extent of the lidar altitude range, but the 12-h wave is present.

WACCM-X shows agreement with wind amplitude seasonal trends, and with phases of the 12-h wave.

Overall, Figure 1 shows that wind amplitudes between lidar, HIAMCM, and WACCM-X appear to be in close agreement, while wind amplitudes are lower in the meteor radar than in the lidar. Furthermore, all four datasets show excellent phase agreement for the 12-h wave above 90 km.

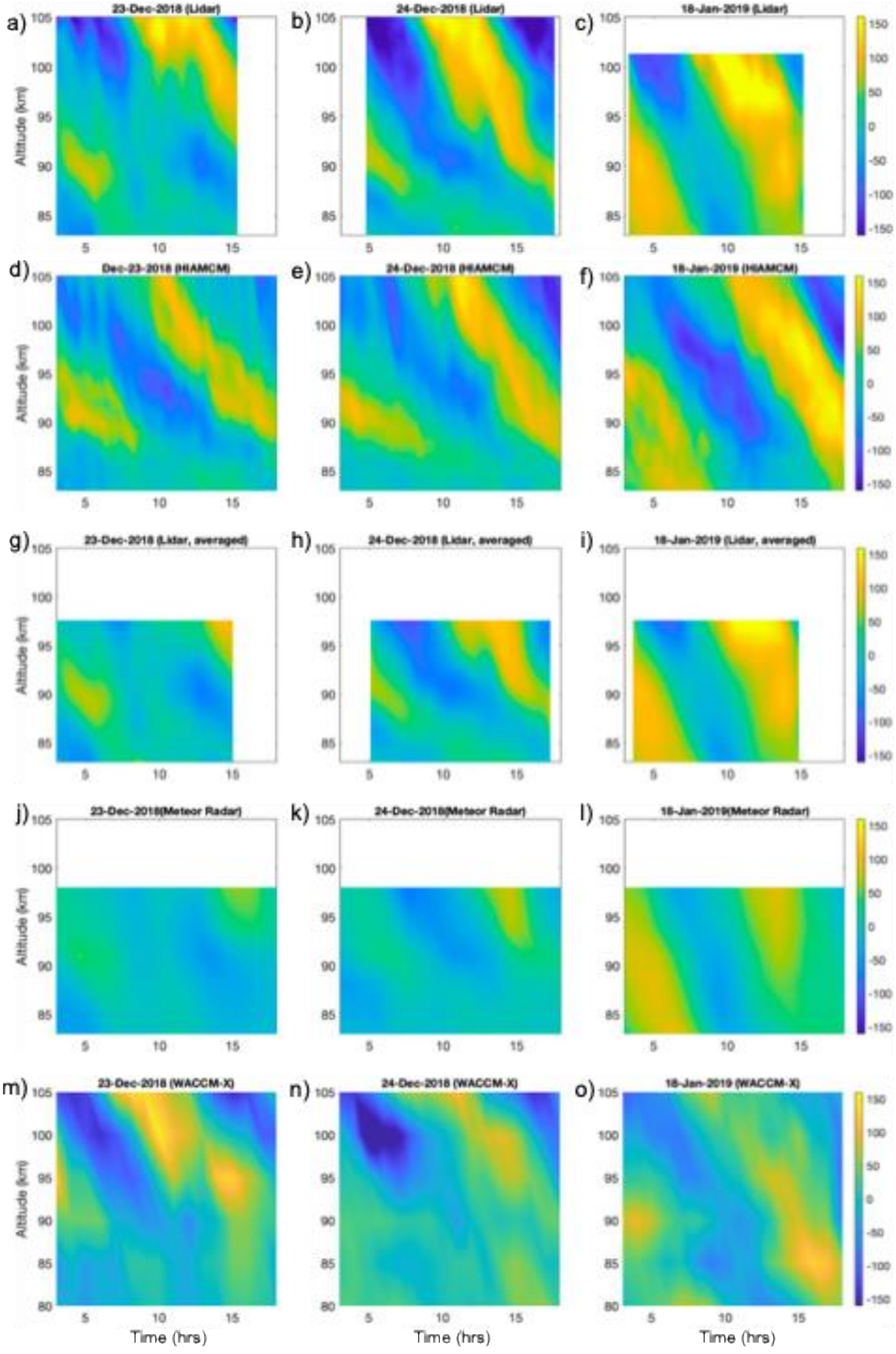
a)
23-Dec-2018 (Lidar)
b)
24-Dec-2018 (Lidar)
c)
18-Jan-2019 (Lidar)
d)
Dec-23-2018 (HIAMCM)
e)
24-Dec-2018 (HIAMCM)
f)
18-Jan-2019 (HIAMCM)
g)
23-Dec-2018 (Lidar, averaged)
h)
24-Dec-2018 (Lidar, averaged)
i)
18-Jan-2019 (Lidar, averaged)
j)
23-Dec-2018(Meteor Radar)
k)
24-Dec-2018(Meteor Radar)
l)
18-Jan-2019(Meteor Radar)
m)
23-Dec-2018 (WACCM-X)
n)
24-Dec-2018 (WACCM-X)
o)
18-Jan-2019 (WACCM-X)
Altitude (km)
Time (hrs)

**Figure 1.** First row is 20-minute temporal resolution, 1-km vertical resolution meridional winds in the lidar (a-c) for a) 23 December 2018, b) 24 December 2018, c) 18 January 2019. The second row shows HIAMCM data for the same nights in panels d-f. The third row is lidar, averaged (g-i) for 1-h and 3-km vertical averaging for the same days. Values are cut off above 98 km for comparison purposes. Third row is meteor radar meridional wind speeds (j-l) for the same days. The final row is WACCM-X (m-o) meridional wind output for the same 3 days. Each row shows different days in the same measurement technique, and each column shows the same day in different measurement techniques.

The 12-h wave is extracted at each altitude of the lidar, meteor radar, WACCM-X, and HIAMCM data using the methods described in Section 2. Figure 2 shows the amplitudes of the fitted 12-h wave at each altitude for the observations and models. The wave amplitudes are shown for 23 December (Figure 2a), 24 December (Figure 2b), 7 January (Figure 2c), and 18 January (Figure 2d).

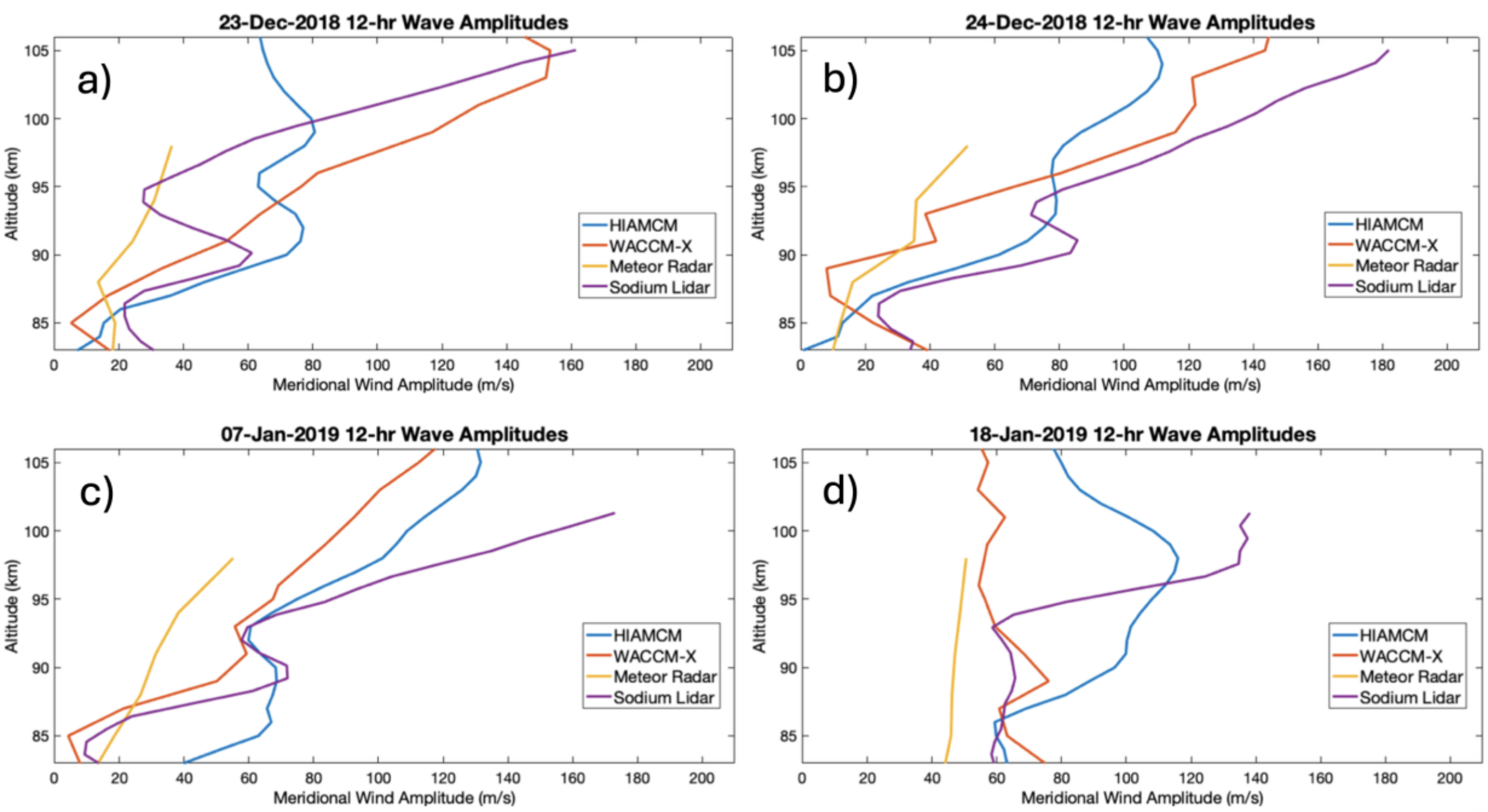

**Figure 2**. Curve fitted 12-h wave fitted at each altitude comparing HIAMCM (blue), WACCM-X (red), meteor radar (yellow), and sodium lidar (purple) for a) 23 December 2018, b) 24 December 2018, c) 7 January 2019, and d) 18 January 2019.

From Figure 2, the sodium lidar shows large 12-h amplitudes, especially at higher altitudes. The sodium lidar measures significant meridional wind amplitudes from 83-105 km that generally grow larger at higher altitudes and can range from 20 m/s to 200 m/s. In December, the 12-h wave amplitudes reached up to 161$\pm$ 3.7 m/s at 105 km on 23 December, and 182 $\pm$ 4.0 m/s at 105 km on 24 December. In January, the amplitudes reached 173 $\pm$ 4.4 m/s at 101 km on 7 January, and 138 $\pm$ 4.2 m/s at 101 km on 18 January.

HIAMCM shows generally similar amplitudes to the lidar at altitudes below 95 km, while lidar amplitudes exceeded HIAMCM amplitudes at altitudes above 95 km. 12-h wave amplitudes in HIAMCM reached 81 m/s at 99 km on 23 December, 112 m/s at 104 km on 24 December, 132 m/s at 105 km on 7 January, and 116 m/s at 98 km on 18 January. HIAMCM agrees with lidar measurements showing 12-h amplitude inversions that occur between 90-95 km on 23 December, 24 December, and 7 January. Differences in amplitude and vertical structure between the lidar and HIAMCM are likely attributable to model shortcomings such as limited resolution, thereby missing some of the GW-tidal interactions.

Overall, WACCM-X shows generally similar 12-h amplitudes to the lidar observed amplitudes, with the exception 18 January above 95 km. In all four nights, lidar measured 12-h wave amplitudes above 100 km exceed HIAMCM and WACCM-X model resolved 12-h wave amplitudes. Furthermore, on all four example nights, the meteor radar shows 12-h wave amplitudes that are significantly lower than the lidar measurements and simulations.

For a seasonal analysis, the 12-h wave amplitude was determined from fits at each altitude for each day of available data in the 2018-2019 season. Only 4 days of HIAMCM data were available for this study, therefore it is not included in our seasonal analysis. Figure 3 shows the meridional amplitudes for the 12-h wave for the lidar, meteor radar, and WACCM-X in panels a-c.

Amplitudes are compared between each dataset by calculating a percent difference, which is represented by

$$\% \text{ Difference} = \frac{\boldsymbol{a_1 - a_2}}{\boldsymbol{(a_1 + a_2)/2}}$$

where $a_1$ and $a_2$ are the 12-h amplitude from two different datasets. This percent difference is calculated at each common altitude between the two datasets, then averaged over all altitudes to yield a single value quantifying their difference.

As mentioned in Section 2, the sodium lidar operated for 13 nights throughout the season. Despite this, the lidar meridional wind data show increases in 12-h wave amplitudes in late December and January that are more than twice the values in November, when amplitude values of less than 60 m/s were observed near 100 km in altitude, while all January dates showed 12-h wave amplitudes between 135-160 m/s near 100 km in altitude. The greatest 12-h wave amplitude in the lidar was calculated to be 209 m/s at 105 km on 7 January. The average error of individual measurements at 105 km on 7 January is $\pm$8.2 m/s, with errors over the night ranging from 0 m/s to 32 m/s for an averaging of 20-minutes and 1-km per data point at that altitude. On this same 7 January date, amplitudes near 100 km in altitude were 173 m/s +/- 3.6 m/s, February dates showed amplitude values of ~75 and ~100 m/s near 100km in altitude, demonstrating a decrease in the 12-h wave amplitude in this time period.

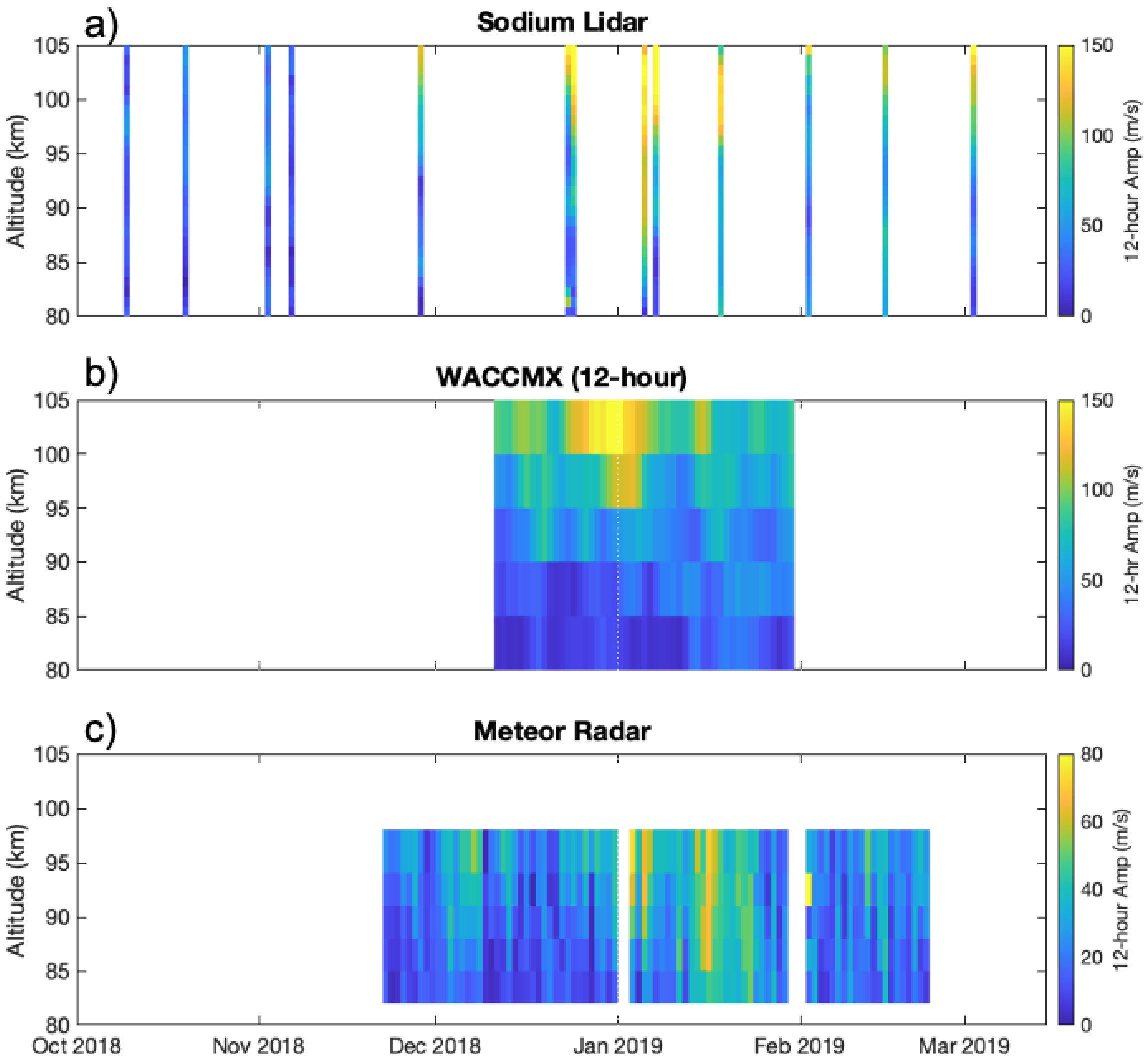

**Figure 3.** a) 12-h wave amplitudes over altitude for available days of data in sodium lidar from 1 October 2018 to 15 March 2019. b) 12-h meridional amplitudes from WACCM-X over altitude (available data between 11 December to January 30) on the same color scale as the sodium lidar (0 to 150 m/s). c) 12-h wave amplitudes over altitude for meteor radar (available data between 22 November to 28 February).

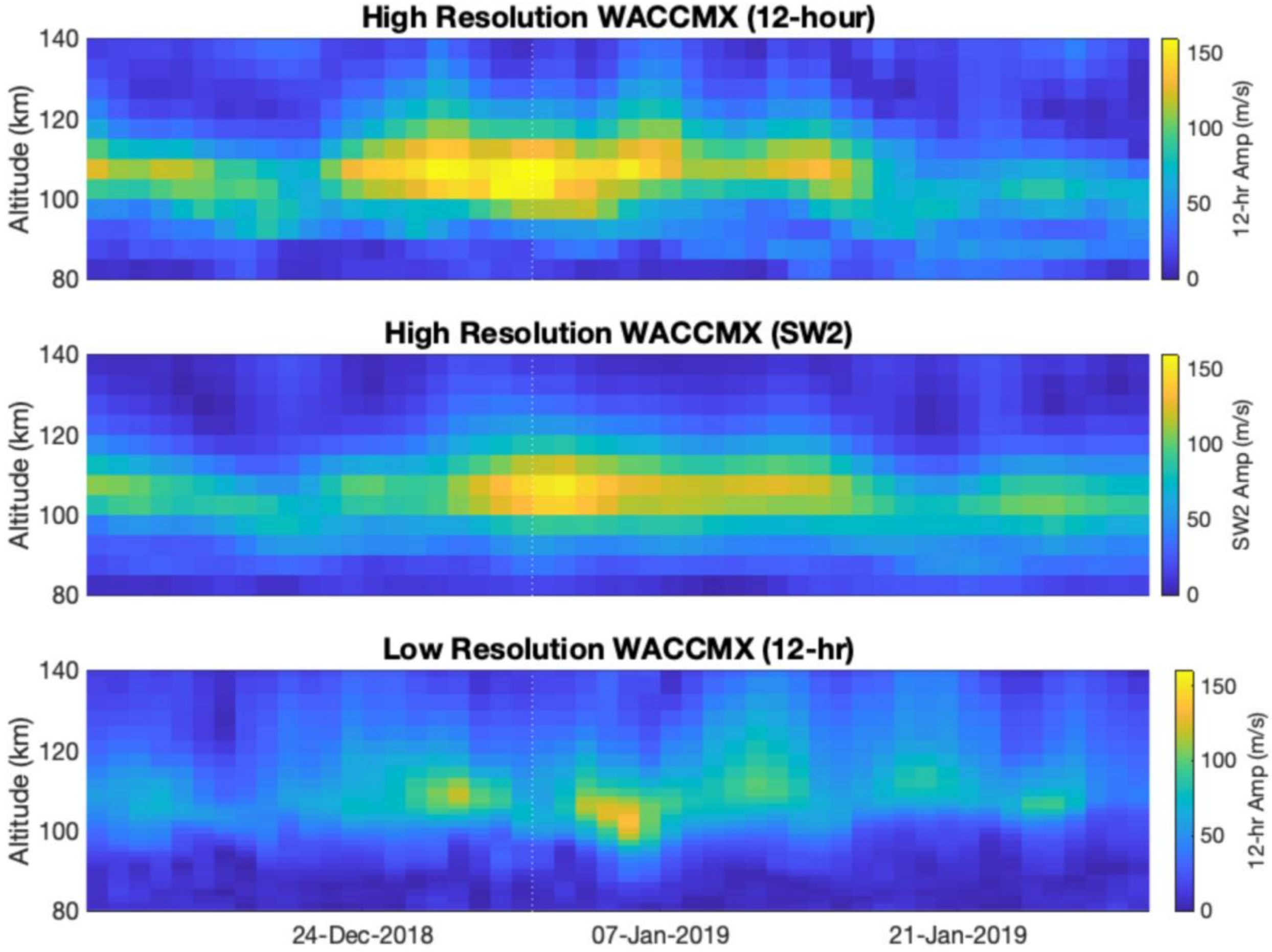

**Figure 4.** a) 12-h wave amplitudes over altitude from WACCM-X and b) globally averaged WACCM-X SW2 amplitudes over altitude from 11 December 2018 to 30 January 2019. Low resolution WACCM-X 12-h wave amplitudes are shown in panel c for the same time.

The meteor radar's continuous dataset allows for a more robust seasonal analysis. The meteor radar did not start operational measurements until late November (22 November 2018), thus data are unavailable before this point. In the radar, 12-h wave meridional amplitudes show some fluctuation in December, minimizing near 9 December and increasing from the end of December to mid-January. Amplitudes then peak between 3 January and 6 January. The largest amplitude measured by the radar, was 75 m/s on 3 January at 98 km.

When constrained to meteor radar altitudes (82-98 km), the largest lidar 12-h amplitude occurred on 5 January with a value of 141$\pm$ 2.0 m/s at 98 km. At this same altitude, the meteor radar fitted 12-h wave amplitude was 73 m/s, showing a 64% difference from the lidar measured amplitude for this day. Wave amplitudes in the meteor radar continuously decrease at all altitudes after this mid-January peak, with a slight increase between 29-30 January, and another slight increase between 9-12 February. Overall, the meteor radar shows similar seasonal trends as the lidar.

When comparing the fitted 12-h wave amplitudes, the percent differences of the amplitudes between 82 to 98 km for lidar and meteor radar are, on average, 58% on 23 December, and 78% on 24 December. In January, the amplitude percent differences are, on average, 59% on 5 January, 44% on 7 January, and 42% on 18 January. This discrepancy may be due to the meteor radar averaging over a large field of view, and the assumption of a uniform wind, $\mathbf{u}=(u,v,w)$ in calculating MLT winds through the radial velocity of every meteor detected in the sky (Hocking et al., 2001).

WACCM-X 12-h meridional amplitudes for this season tend to peak between 100 km and 115 km, remaining fairly consistent around 100 m/s before the SSW. On 23 December, amplitudes begin to steadily increase through 8 January, maximizing on 1 January with a value of 162 m/s at 105 km. For 23 December, the percent difference within 85-105 km between the lidar and WACCM-X data ranges from -84% at 95 km (the negative indicates WACCM-X reports larger amplitudes than the lidar), and 69% at 85 km, giving an average percent difference of 9% (this low number is misleading due to the large range in percent difference values). On 24 December, the percent difference ranges from 11% at 95 km to 112% at 85 km, with an average percent difference of 47%.The largest percent differences between WACCM-X and lidar arise on 5 January. On this day, the percent difference between 85-105 km ranges from -11% at 105 km to 127% at 85 km, with an average percent difference of 47%.

The largest amplitudes of the season for the WACCM-X run occur on 1 January, reaching a maximum amplitude value of 163 m/s at 105 km altitude. There is no lidar data to compare for this day, so we compare the model to lidar for the day with the largest lidar amplitudes, 7 January. The largest amplitudes for the WACCM-X on 7 January reach a maximum amplitude value of 130 m/s at 105 km altitude. WACCM-X's maximum 12-h amplitude is 47% different than the lidar's calculated 209 m/s at 105 km on the same date.

After 8 January, 12-h wave amplitudes remain lower until a smaller increase from 12-16 of January, before decreasing through the end of January. We noted in Figure 1 a larger difference in wind amplitudes between lidar and WACCM-X data for 18 January. For this day's 12-h wave, the percent difference within 85-105 km ranges from 12% at 90 km, and 66% at 100 km, giving an average percent difference of 29%.

Figure 4 compares WACCM-X 12-h wave amplitudes to SW2 amplitudes at the latitude of Chatanika for the season. The seasonal variation observed in WACCM-X SW2 amplitudes is very similar to that seen in the WACCM-X 12-h wave amplitudes, although the amplitude values differ; SW2 meridional amplitudes are slightly lower than the 12-h wave amplitudes over Alaska. The average 12-h amplitude from 23 December to 16 January at 105 km was 139 m/s, whereas the average SW2 amplitude for the same time period was 122 m/s – 13% different. The seasonal maximum SW2 amplitude occurs very slightly later than the maximum in the localized 12-h wave over PFRR, peaking between 1-3 January with slightly lower amplitudes between 144-153 m/s at 105 km. Both the 12-h wave and tidal amplitudes slightly decrease but remain elevated through the middle of January, with a decrease after this time. The comparison here indicates that the 12-h wave amplitudes observed by the meteor radar and lidar are a good tracer for overall trends in SW2 activity, but slightly overestimate the amplitudes on a global scale.

From these results, we find that the general seasonal trend is consistent amongst the lidar, radar, and WACCM-X. As seen in Figure 1, the meteor radar shows consistently lower 12-h wave amplitudes than the lidar measurements. Both models, HIAMCM and WACCM-X, do well in capturing MLT wind amplitudes when compared to the lidar, leading to fair estimations of the 12-h wave. Overall percent differences between lidar and WACCM-X range from 8-47%.

## 4 Discussion

These lidar observations indicate that the 12-h wave amplitude, a proxy for the SW2 tide, may be larger than previously reported in the Arctic region. The lidar measurements yield consistently larger amplitudes than the meteor radar measurements; this follows a previous study focused over New Mexico that showed lidar observed winds at 93 km were larger than the meteor radar observations (Table 2 in Liu et al., 2002). Given that the HIAMCM and WACCM-X simulations of the 12-h wave show reasonable agreement in amplitude with the lidar, the local amplitude of SW2 is likely larger than previously recognized. Previous HIAMCM-based studies have shown higher magnitudes than the meteor radar (Figure 19 in Vadas et al., 2024). The large 12-h wave amplitudes may reflect the impact of secondary GWs (Hindley et al., 2022) or GW-tidal interactions (Becker & Oberheide, 2023). Large amplitude 12-h waves at the South Pole have been attributed to the modulation of GWs by the semidiurnal tide (Collins et al., 1992). If GW-tidal interactions are important for the tidal components in the winter MLT, shortcomings in the simulation of secondary GWs could be a source of uncertainty, and therefore would increase the discrepancy between model runs and observations.

The high-resolution WACCM-X results presented in this study show far larger amplitudes, similar to lidar measurements, than low-resolution WACCM-X, as shown in Figure 4c. This is an improvement with respect to the low-resolution model that has been used in previous studies. The low-resolution WACCM-X run does not parameterize GWs above ~100 km; this indicates that either GW-tide interactions or background winds (which are better represented in high-resolution WACCM-X) are significantly impacting the SW2 and 12-h wave amplitudes in the high-resolution WACCM-X. The high-resolution WACCM-X resolves GW dynamics, which allows for more accurate GW dissipation impacts on tides that should give rise to larger tidal amplitudes (Liu et al., 2008). The effect of GW drag on tidal amplitudes is also supported by HIAMCM's meridional winds being similar to the lidar values; HIAMCM resolves primary and higher-order GWs ($\boldsymbol{\lambda_H > 200}$ km) created by GW dissipation (Becker et al., 2022; Vadas et al., 2024). However, HIAMCM shows lower meridional wind values than the sodium lidar in Figure 1 at altitudes above 95 km.

The higher amplitude of the observed 12-h wave may also be due to the local superposition of the migrating (SW2) and non-migrating (SE3) tides. The WACCM-X simulation allows characterization of the SW2 and SE3 non-migrating tidal components (Figure 5). The amplitude of the 12-h non-migrating tide increases in January, with a maximum amplitude of 69 m/s at 105 km on 5 January. While the SE3 amplitudes are half the magnitude of the SW2 at the same day/altitude, they are significant enough to contribute to the difference in localized 12-h wave amplitudes over Chatanika, AK and globally measured SW2 tidal amplitudes. This would imply that the migrating SW2 tides are being accurately characterized in the models, and the discrepancies reflect limitations in simulating the amplitude and/or phase of the nonmigrating semidiurnal tides.

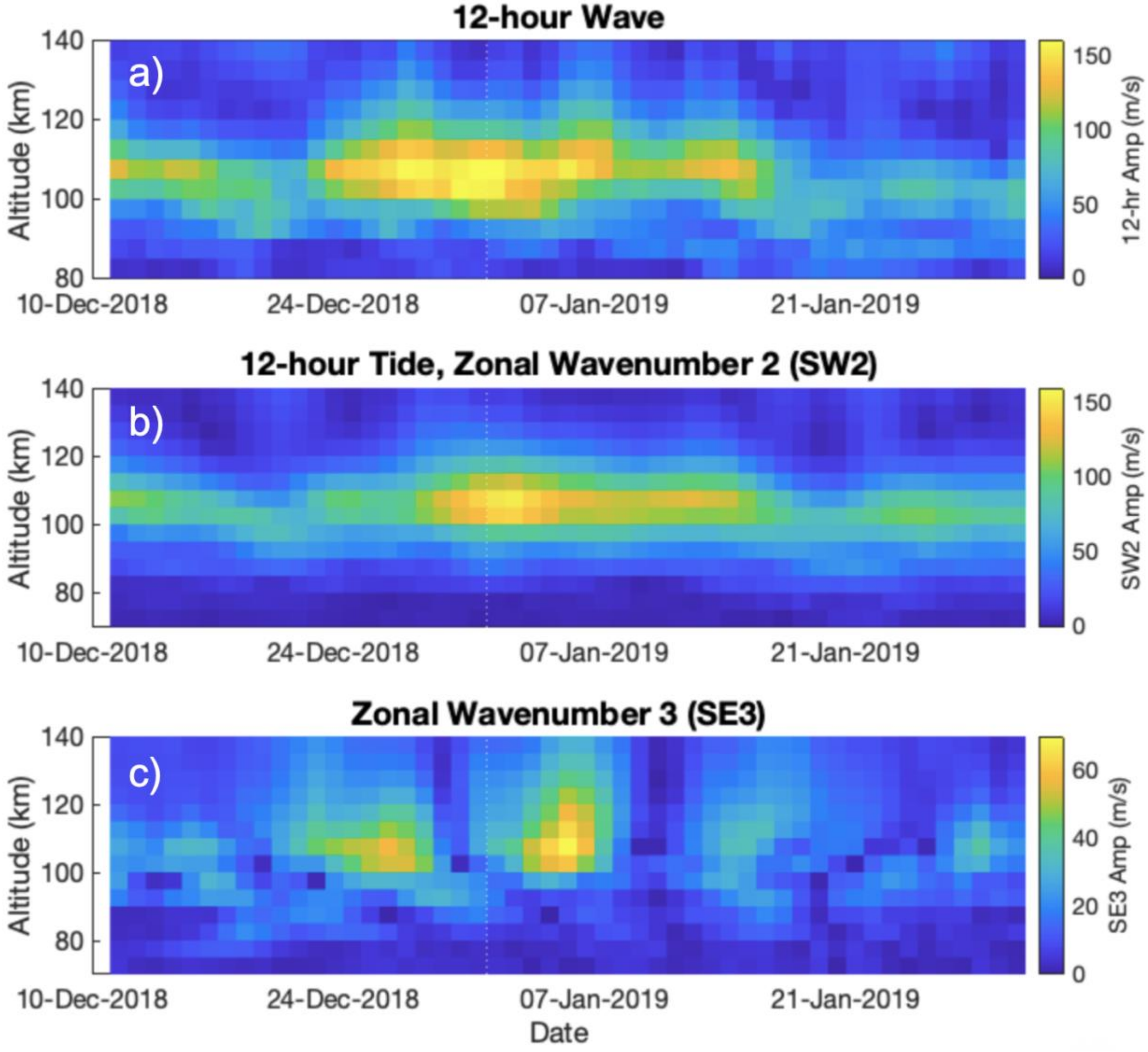

**Figure 5**. WACCM-X's a) 12-h wave amplitudes, over Poker Flat, Alaska, b) SW2 tidal amplitudes, and c) nonmigrating SE3 tidal amplitude, all from 11 December to 30 January. A white dashed line denotes 1 January 2019.

We consider the relationship between the 12-h wave amplitude and both vortex strength and geomagnetic activity. Figure 6a presents the 12-h wave's meridional amplitudes at 95 km for the lidar (black asterisks), radar (blue line), and WACCM-X (yellow line) with the polar vortex strength (purple line) and daily average SME index (grey line) in Figure 6b. As previously shown in Figure 1, the sodium lidar measurements typically reveal higher 12-h wave amplitudes than those measured by the meteor radar (64% difference at season peak). For corresponding days in December and January, the percent difference in 12-h wave amplitudes between lidar and WACCM-X was -84% (23 December), 11% (24 December), 37% (5 January), 33% (7 January), and 11% (18 January). The large discrepancy on 23 December where the lidar 12-h amplitude at 95 km (27.9 m/s) appears similar, if not far less than, both the meteor radar (30.9 m/s) and WACCM-X (68 m/s) 12-h amplitude. This low lidar measurement coincides with the presence of a large amount of GW variation in the residual measurement after the subtraction of the 12-h wave. These GWs may result in biasing of the 12-h fit and/or cause a GW-tidal interaction that lowers the tidal amplitude. This lower 12-h amplitude fit at 95 km coincides with large 12-h wave amplitudes above 95 km exceeding 100 m/s (Figure 2d).

Regardless of the differences between the specific wind values, the lidar, radar and model all show enhanced 12-h wave amplitudes during the SSW period (25 December through 10 January) when zonal winds at 50 km reversed from eastward to westward (Phillips et al., 2026). Studies of ICON and COSMIC-2 satellite-based measurements have shown SW2 amplitudes to increase by up to a factor of two during SSWs (Oberheide, 2022). Scanning Doppler Imager measurements of 12-h wave amplitudes in the F region over Chatanika have reported increases in amplitude of up to 100% (Kumari et al., 2025).

The Northern Annular Mode (NAM) provides a measurement of the strength of the polar stratospheric vortex (Gerber & Martineau, 2018). The NAM index at 10 hPa is used to represent polar vortex strength. Values less than -2 represent a weak vortex state and indicate the vortex breakdown (Figure 6b). The effect of the SSW is clearly visible in early January when the NAM index remains low (-3) over several weeks from late December to mid-January.

The Global SuperMAG auroral Electrojet (SME) index averages the measurements from 100 magnetometers at high latitudes every 1-minute to quantify global auroral activity (Newell & Gjerloev, 2014; Gjerloev, 2012). The SME index is plotted for comparison to 12-h wave amplitudes and shown in Figure 6b. Corresponding lidar days with SME index are indicated by red dots. Although the Arctic winter 2018-2019 occurred during solar minimum, limiting the solar influence on the upper atmosphere and MLT compared to solar maximum, there were still some days that experienced enhanced geomagnetic activity. Still, there is no obvious variation seen between the two measurements. For one day, 5 January, we see a very large 12-h wave amplitude in the lidar (118 m/s) and in the WACCM-X (79.8 m/s), and this day does indeed align with a peak in geomagnetic activity (SME=290 nT). Yet on days with larger geomagnetic activity, such as 9 October (SME=462 nT) or 2 March (SME=343 nT), the lidar amplitudes (31.4 m/s and 71.3 m/s, respectively) are far less than those on 5 January.

The meteor radar 12-h amplitudes at 95 km and SME index also do not appear to have a clear connection. We note that there are specific dates when localized amplitude peaks align with peaks in the SME index, similar to the lidar. This includes 5 January, where the radar showed a 12-h amplitude of 63.8 m/s.

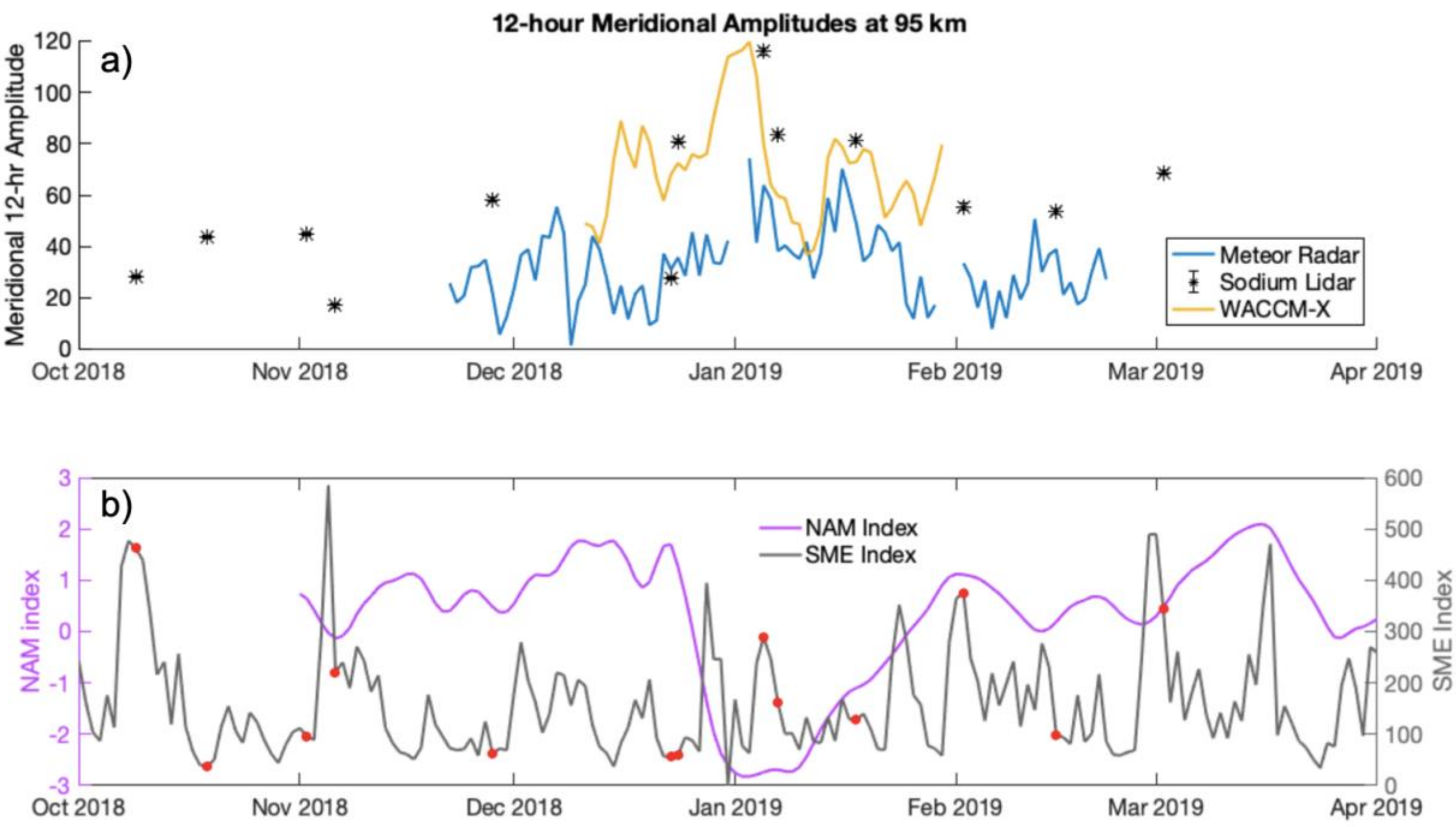

**Figure 6**. a) 12-h fit meridional wind amplitudes at 95 km for sodium lidar (black), meteor radar (blue), and WACCM-X (yellow) from 1 October 2018 to 1 April 2019. b) SME index (grey) and NAM index (purple) for the same time period. Red dots indicate days coinciding with lidar measurements.

## 5 Conclusions

In the presented research findings, 12-h wave meridional amplitudes over Alaska were observed by sodium lidar measurements during the 2018-2019 Arctic winter, with maxima's ranging between 138 m/s and 182 m/s in December and January. These amplitudes were compared to the meteor radar observations, and model runs during the same time frame for both the 12-h wave and SW2 amplitudes. Conclusions are summarized below:

- The sodium lidar meridional wind measurements of the 12-h wave show significantly higher amplitudes than previous meteor radar measurements in the polar region, often exceeding meteor radar amplitudes by a factor of two at altitudes above 95 km.
- HIAMCM presented meridional wind 12-h wave amplitudes that were also larger than those from the meteor radar, but similar to those from the sodium lidar at altitudes below 95 km. However, these HIAMCM amplitudes were smaller than the sodium lidar measurements at altitudes above 95 km.
- WACCM-X reproduces the overall 12-h wave trends observed in the lidar and meteor radar for the season, showing a peak in amplitude in early January. This high-resolution model's ability to capture GW-tidal interactions substantially improves the agreement between the observed and modeled amplitudes of the 12-h wave compared to earlier low-resolution version of WACCM-X.
- WACCM-X shows comparable but typically smaller 12-h wave meridional amplitudes than those measured by the sodium lidar.
- The meteor radar shows smaller 12-h wave amplitudes than the sodium lidar by up to 54% at 95 km during peak amplitude times in January. This is likely due to the resolution of both observational methods, as the radar averages data over 1 hour and 3 km covering 170 degrees of the sky, whereas the lidar processed here has a higher temporal and altitude resolution with a footprint covering less than 100 m.
- A clear, consistent relationship with the 12-h wave meridional amplitudes and geomagnetic activity was not found. However, it was observed that the largest amplitudes occur during the SSW, when the polar vortex strength is at its minimum.

The observations suggest that, based on lidar meridional wind measurements, high-latitude 12-h wave amplitudes may be significantly larger than previously reported by meteor radars.

WACCM-X does well in showing the seasonal trend of the 12-h wave when compared to both sets of observations. Lidar comparison with both HIAMCM and WACCM-X demonstrated good agreement at altitudes less than 95 km, but indicates that 12-h wave amplitudes may be larger than modeled at altitudes between 95-105 km.

## Acknowledgments

SRP and KB acknowledge funding from NASA FINESST Grant 80NSSC24K1865. NP acknowledges support from NASA grants 80NSSC24K0269 and 80NSSC23K0418. EB acknowledges funding from NSF grant 2329957 and NASA grant 80NSSC24K0274. RC acknowledges support from the NSF under grant AGS 1829161 for the lidar operations and analysis. We acknowledge the staff at PFRR for their continued support of the observational programs and Jennifer Alspach and Jintai Li for their assistance with the lidar observations. We thank Biff Williams for providing the raw sodium lidar data which can be publicly accessed at https://boulder.gats-inc.com/~biff/PokerFlatNaLIdar/index.html. We acknowledge support from the National Center for Atmospheric Research, which is a major facility sponsored by the U.S. National Science Foundation under Cooperative Agreement 1852977. We acknowledge the contributions of the SuperMAG collaborators https://supermag.jhuapl.edu/info/?page=acknowledgement.

## Open Research

Sodium lidar data is available at https://boulder.gats-inc.com/~biff/PokerFlatNaLIdar/index.html. HIAMCM outputs are also available online through the corresponding author's Zenodo (Phillips, 2026). Meteor radar winds are located on Zenodo and were originally published by Kumari (2024). WACCM-X is run through the National Center for Atmospheric Research; the model setup code is available through https://zenodo.org/records/10895515 [Software]. NAM index is calculated using MERRA-2 data; MERRA-2 wind data is publicly available after free registration to the NASA GES DISC page at https://goldsmr5.gesdisc.eosdis.nasa.gov/data/MERRA2/M2I3NPASM.5.12.4/ . MERRA-2 winds are accessible after free registration to the NASA GES DISC webpage (Global Modeling and Assimilation Office (GMAO), 2015). SME index was obtained through public download at https://supermag.jhuapl.edu/.

## Conflict of Interest Disclosure

The authors declare there are no conflicts of interest for this manuscript.